\documentstyle[aps,graphicx]{revtex}
\title{Conductivity and Thickness of DNAs} 

\author{A.Yu. Kasumov$^{1,2}$\thanks{
Present address: RIKEN, Wako, Saitama, 351-0198 Japan. E-mail: kasumov@postman.riken.go.jp. 
To whom correspondence should be addressed.}, D.V. Klinov$^3$} 

\address{
$^1$Laboratoire de Physique des Solides, Associ\'e au CNRS, B\^at 510,
Universit\'e Paris--Sud, 91405, Orsay, France. 
$^2$Institute of Microelectronics Technology and High Purity Materials, 
Russian Academy of Sciences, Chernogolovka 142432 Moscow Region, Russia. 
$^3$Shemyakin-Ovchinnikov Institute of Bioorganic Chemistry, Russian 
Academy of Sciences, Miklukho-Maklaya 16/10, Moscow 117871, Russia.}

\begin{document}
\maketitle

\relax 
\citation{7}
\citation{4}
\citation{4}
\citation{8}
\citation{9}
\citation{10}
\citation{1}
\citation{2,3,5}
\citation{1}
\citation{6}
\citation{11}
\citation{4}
\bibcite{1}{1}
\bibcite{2}{2}
\bibcite{3}{3}
\bibcite{4}{4}
\bibcite{5}{5}
\bibcite{6}{6}
\bibcite{7}{7}
\bibcite{8}{8}
\bibcite{9}{9}
\bibcite{10}{10}
\bibcite{11}{11}

\begin{abstract}
Debates about conductivity of DNAs have been 
recently renewed due to contradictory results of direct measurements by use 
of electrical contacts to molecules. In several works it was discovered that 
double-stranded (ds)DNAs are conductors: metals or semiconductors [1-6]. 
However in other works \cite{7} the absence of DNAs conductivity has been 
observed even for the molecules with ordered base pairs structure. Here we 
show that the absence of conductivity is caused by a very large compression 
deformation of DNAs. Thickness of such compressed DNAs is 2-4 times less 
than the diameter (about 2nm) of native Watson-Crick B-DNA.
\end{abstract}

$\lambda - $DNA molecules were deposited (from the same buffer solution as in 
\cite{4}) on mica substrates partially covered by a Pt film with thickness of 3 
nm. Using an AFM microscope operating both in standard and spreading 
resistance (SRM) modes it was possible to measure simultaneously the height 
and conductivity of the same molecules, crossing the edge of the Pt film.

In the absence of any treatment of the mica+Pt substrate we observed the 
DNAs with typical height about 1 nm and no contrast in the SRM mode 
indicating insulating molecules in agreement with previous observations 
(Fig. 1a,b). On the other hand, very different results are obtained if prior 
to deposition of molecules, a thin (about 0.5 nm) layer of island polymer 
film is sputtered on the surface of both Pt and mica by glow discharge of 
pentylamine vapor, as it was done in our previous experiments \cite{4}. The 
thickness of observed DNA molecules was about 2 nm and they were clearly 
visible in SRM (Fig.1c,d). We interpret this native thickness the following 
way: the deposition of the polymer film decreases hydrofilicity of mica and 
thus its interaction with DNAs. Average thickness of DNA molecules on the 
substrates treated by pentylamine is 2.4 $\pm $ 0.5 nm for 64 measurements 
on different molecules; for DNAs on the clean substrate this value is 1.1 
$\pm $ 0.2 nm for 57 measurements.

Careful studies in AFM have shown that hugely reduced thickness of DNAs on 
the clean mica and silicon substrates is real, but not an artifact of 
microscopy \cite{8}. We additionally checked it by transmission electron 
microscopy replica method without use of AFM. DNAs on the clean surface have 
the same contrast as mica, and so they are insulators. On the Pt surface 
some of DNAs are seen in negative contrast (Fig.1b). Such a contrast was 
observed by STM and explained by insulating behavior of DNAs previously \cite{9}. 
Contrariwise DNAs on the mica covered by polymer film are visible by SRM in 
positive contrast (Fig 1d,f). Thus they are conductors.

We believe the conductivity of DNAs comes from the native, periodic 
structure of the molecules. It is well known that the mechanical stretching 
deformation can lead to denaturation of DNAs \cite{10}. Likewise, compressing
deformation may be able to destroy the periodic structure of DNAs. 
Compressed DNA with thickness 2 - 4 times less than B-DNA represent two 
independent, chaotically intersecting strands. Single-stranded DNAs 
definitely are insulators \cite{1}.

Brief review of works [1-6] shows dsDNAs are conductors if they have native 
thickness about 2nm (in this case we can call it "diameter"). For suspended 
DNAs \cite{2,3,5} and DNA films \cite{1} there is no interaction with a surface, and 
the thickness should be close to native. For thick ropes of DNAs \cite{6} the 
interaction is weak for internal DNAs inside a rope. Conductivity of native 
DNAs of course will depend on many factors, such as the contacts to the 
electrodes \cite{11}. DNA will behave like a metal if the contacts are ohmic and 
the distance between them is less than a screening length of the molecule. 
Poor screening is a well known property of 1D systems \cite{4}. It seems to be 
true for DNAs between Re/C contacts (Fig.1e) with resistance about 100 kOhms 
per molecule. Different types of kinks and bendings can also decrease 
conductivity of DNAs. Straight DNAs (Fig.1e,f) are more conductive than 
curved ones (Fig.1d). From the above results we conclude that dsDNA 
molecules can be conductors, and one can use them in molecular electronics.

We acknowledge fruitful discussions with H.Bouchiat, V.Croquette and 
D.Bensimon. A.K. thanks the Russian Foundation for Basic Research and Solid 
State Nanostructures for financial support and thanks CNRS for a visitor's 
position.

\begin{center}
\begin{figure}
\includegraphics[width=12cm]{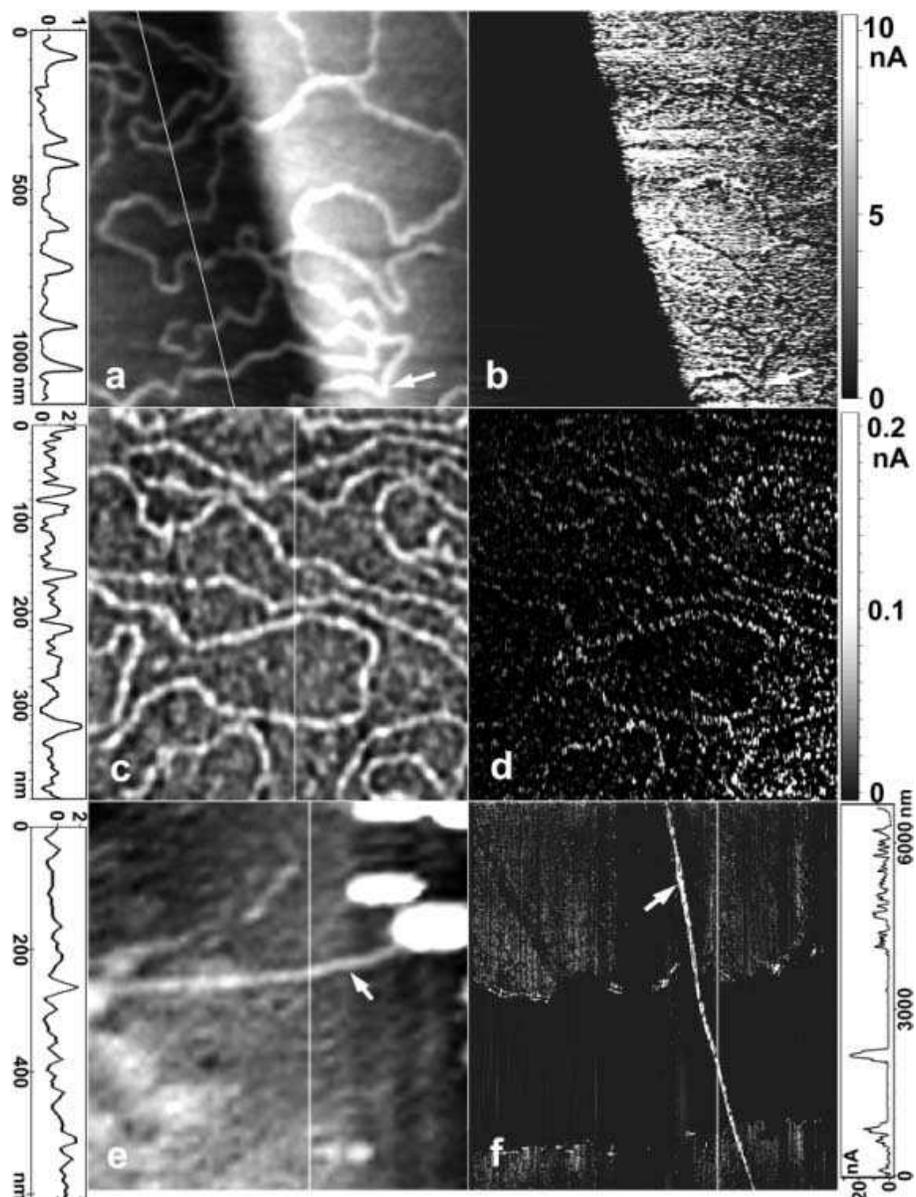}
\caption{
AFM (left) and SRM (right) pictures of 
DNAs: (\textbf{a}) AFM picture of DNAs on the clean substrate without 
pentylamine; (\textbf{b}) SRM picture of the same molecules (right bright 
part of a and b pictures is Pt); (\textbf{c}) AFM picture of DNAs on the 
substrate treated by pentylamine; (\textbf{d}) SRM picture of the same 
molecules, Pt electrode is outside of the picture; (\textbf{e}) AFM picture 
of a DNA combed (as in [4]) across the slit between Re/C electrodes on mica; 
(\textbf{f}) SRM picture of a rope of DNAs combed (as in [4]) across the 
slit between Pt electrodes on mica. Some of DNAs are shown by arrows. From 
the left and right sides of the picture there are profiles of DNAs (height in nm)
and current scales of SRM pictures (voltage was up to 0.23 V) respectively.
}
\end{figure}
\end{center}

\end{document}